\documentstyle[psfig,prd,aps]{revtex}
\begin{document}
\draft

\title{Choptuik scaling in six dimensions}

\author{David Garfinkle
\thanks{Email: garfinkl@oakland.edu}}
\address{
\centerline{Department of Physics, Oakland University,
Rochester, Michigan, 48309}} 

\author{Curt Cutler
\thanks {Email: cutler@aei-potsdam.mpg.de}}
\address{
\centerline{Max Planck Institute for Gravitational Physics, D-14476 Golm, 
Germany}} 

\author{G. Comer Duncan
\thanks {Email: gcd@chandra.bgsu.edu}}
\address{
\centerline{Department of Physics and Astronomy, Bowling Green State University,
Bowling Green, Ohio 43403}}

\maketitle

\null\vspace{-1.75mm}

\begin{abstract}

We perform numerical simulations of the critical gravitational collapse of a 
spherically symmetric scalar field in 6 dimensions.  The critical
solution has discrete self-similarity.  We find the critical exponent $\gamma$
and the self-similarity period $\Delta$.
 
\end{abstract}


\section{Introduction}

Critical behavior in gravitational collapse, 
as first found by Choptuik\cite{choptuik},
occurs at and near the threshold of black hole formation\cite{carstenr}. 
For a one-parameter family of initial data 
slightly above the threshold, the mass of the 
black hole, $M_{BH}$, scales like ${(p-p*)}^\gamma$.  
Here $p$ is the parameter, $p*$ is its
critical value and $\gamma$ is a constant that depends on the type of 
matter, but not
on the family of data. 
For initial data slightly below the threshold, the maximum
curvature scales like ${(p*-p)}^{- 2 \gamma}$ where $\gamma$ is 
the same constant as
in the black hole mass scaling law\cite{meandcomer}. 
(The definition of the scaling  exponent $\gamma$ must be 
slightly generalized in $n$ spacetime 
dimensions, where $M_{BH}$ has dimension $(\rm length)^{n-3}$; see section II.) 
The critical solution ($p=p*$)
has either continuous self similarity or discrete self similarity, 
depending on the type
of matter.

While critical gravitational collapse has been studied in many types of matter, 
the work has, in general, been done in 4 spacetime dimensions.  
(The exception is
work on analogs of gravitational collapse in 3 spacetime dimensions\cite{alan}).
One might therefore wonder whether critical behavior occurs in 
gravitational collapse
in $n$ spacetime dimensions for $n>4$, and if so, how the properties 
of the critical
behavior depend on $n$.

In this work, we perform numerical simulations of the collapse of a spherically
symmetric scalar field in 6 spacetime dimensions.  We find that
the critical solution has discrete self-similarity.  We find the 
scaling exponent
$\gamma$ and the self-similarity period $\Delta$.  
Section 2 reviews the Schwarzschild solution in $n$ dimensions and shows how 
the definition of the $\gamma$ exponent should be generalized to the 
$n$-dimensional case.
Section 3 gives
the equations for the evolution of the scalar field and the metric in a form 
suitable for our numerical simulations.  The numerical method is presented in
section 4 and our results in section 5.

\section{Schwarzschild in $n$ dimensions}
The Schwarzschild metric in $n$ spacetime dimensions is \cite{myers_perry}
\begin{equation}
\label{schw}
ds^2 = -\bigl(1-\frac{16\pi}{(n-2)A_{n-2}}\frac{M}{r^{n-3}}\bigr)dt^2 +
\bigl(1-\frac{16\pi}{(n-2)A_{n-2}}\frac{M}{r^{n-3}}\bigr)^{-1} dr^2 + r^2 
d\Omega^2_{n-2}
\end{equation}
\noindent where $A_{n-2} = 2\pi^{(n-1)/2}/\Gamma(\frac{n-1}{2})$ is the
area of the unit $(n-2)$-sphere. The quantity $M$ in the metric (\ref{schw})
is the spacetime mass. M is normalized so that when (\ref{schw})
is the exterior metric of a nearly Newtonian static fluid ball, then
$M$ is $\int{T_{00} dV}$. With this normalization, the force on
a static unit-mass particle at distance $r$ approaches
$\frac{8\pi (n-3)}{(n-2)A_{n-2}} M/r^{n-4}$ as $r \rightarrow \infty$.   

It is important to note that $M$ has dimension $({\rm length})^{n-3}$.
This implies the following scaling behaviors for critical 
phenomena in $n$ dimensions. If below threshold the maximum curvature 
[which has dimension $({\rm length})^{-2}$]
scales like $(p* - p)^{-2\gamma}$, then above threshold $M_{BH}$ 
scales as $(p - p*)^{(n-3)\gamma}$.

\section{Evolution equations}

Our matter model is the one used in reference\cite{choptuik}: a massless, 
minimally coupled, self gravitating scalar field
$\phi$. The stress-energy of the field is
\begin{equation}
\label{se}
{T_{ab}} = {\nabla _a} \phi \, {\nabla _b} \phi \; - \; {1 \over 2} \; 
{\nabla ^c} \phi {\nabla _c} \phi \, {g_{ab}} 
\end{equation}
In $n$ spacetime dimensions, a spherically symmetric metric has the form
\begin{equation}
d {s^2} = - \; {e^{ 2 \nu}} \, d {u^2} \; - \; 2 \, {e^{\nu + \lambda }} 
\, d u \, d r \; + \; {r^2} d {\Sigma ^2}
\end{equation}
where
$ d {\Sigma ^2} $ is the metric of the unit
$ (n - 2)$-sphere and $ \nu $ and $ \lambda $ are functions of 
$ u $ and $ r $.  The coordinate $r$ is a generalization of the 
usual area radius.
The coordinate $u$ is constant on outgoing radial null geodesics.  
Here, we also
require that on the world line of the central observer, $u$ is equal 
to the proper
time of that observer.  This gives rise to the condition $\nu = 0 $ at $r=0$.
Define $ {E_{ab}} \equiv {G_{ab}} \, - \, 8 \pi {T_{ab}} $ where $G_{ab}$ is the
Einstein tensor, so the Einstein-scalar equations are just
$ {E_{ab}} = 0 $.  Note that as a consequence of the Bianchi identities, $\phi$
must satisfy the wave equation.  

In 4 spacetime dimensions, 
Christodoulou\cite{christodoulou} showed
how to write the spherically symmetric Einstein-scalar equations as an
integro-differential equation for the scalar field.  We now generalize 
the method of
reference\cite{christodoulou} to the $n$-dimensional case. 
Define the null vectors
$ l^a $ and $ n^a $
by
\begin{equation}
\label{ldef}
{l^a} = {e^{ - \lambda }} \; {{\left ( {\partial \over {\partial r}} \right
) }^a} 
\end{equation}
\begin{equation}
\label{ndef}
{n^a} = {e^{ - \nu }} \; {{\left ( {\partial \over {\partial u}} \right )
}^a}
\; -
\; {1
\over 2} \; {e^{ - \lambda }} \; {{\left ( {\partial \over {\partial r}} \right ) }^a} 
\end{equation}
Then some straightforward but tedious calculations give
\begin{equation}
\label{gll}
{G_{ab}} {l^a} {l^b} = {{n - 2} \over r} \; {e^{ - 2 \lambda }} \;
{\partial
\over {\partial r}} \; ( \lambda \, + \, \nu ) \; \; \; ,
\end{equation}
\begin{equation}
\label{gln}
{G_{ab}} {l^a} {n^b} = {{n - 2} \over 2} \; \left [ {{n - 3} \over {r^2}}
\; \left ( 1
\, - \, {e^{ - 2 \lambda }} \right ) \; - \; {{e^{ - 2 \lambda }} \over r} \; {\partial
\over {\partial r}} \; ( \nu \, - \, \lambda ) \right ] \; \; \; .
\end{equation}
One can show that the equations
\begin{equation}
\label{Eequiv}
{E_{ab}} {l^a} {l^b} = 0 \; \; \; \; , \; \; {E_{ab}} {l^a} {n^b} = 0 \; \; \; \;
, \; \; \; {\nabla _a} {\nabla ^a} \phi = 0 
\end{equation}
plus the condition of regularity at the origin imply
$ {E_{ab}} = 0 $ everywhere.
Thus we need to impose only those three equations.  

Using equations (\ref{gll},\ref{gln}) and (\ref{se}), we find that the first two of
equations (\ref{Eequiv}) yield
\begin{equation}
\label{gllc}
{{n - 2} \over r} \; {\partial \over {\partial r}}
\; (
\lambda
\, + \, \nu ) \; - \; 8 \pi \; {{\left ( {{\partial \phi } \over
{\partial r}} \right ) }^2} = 0 \; \; \; .
\end{equation}
\begin{equation}
\label{glnc}
{{2 - n} \over {2r}} \; \left [ 
\; e^{ - 2 \lambda }\; {\partial \over {\partial r}}\; (
\nu \, - \, \lambda ) 
- 
{{n - 3} \over {r} } \; \left ( 1 \, - \,{e^{ - 2\lambda }} \right ) \; 
\right ] = 0 \; \; \; .
\end{equation}
Define $g \equiv {e^{\nu + \lambda}}$ and ${\bar g} \equiv {e^{\nu - \lambda}}$.
Then the solution of equations (\ref{gllc}) and (\ref{glnc}) is 
\begin{equation}
\label{gsol}
g = \exp \left [ {{8 \pi } \over {n - 2}} \; {\int _0 ^r} \; {\hat r} \;
{{\left ( {{\partial \phi } \over {\partial {\hat r}}} \right ) }^2} \; d {\hat r} \right ] \; \; \; .
\end{equation}
\begin{equation}
\label{gbarsol}
{\bar g} = {{n - 3} \over {r^{n - 3}}} \;  {\int _0 ^r} \;
{{\hat r}^{n - 4}}
\; g(\hat r) \; d \hat r \; \; \; .
\end{equation}
The wave equation for 
$ \phi $
in this metric is
\begin{equation}
\label{wave}
2 \; {{{\partial ^2} \phi } \over {\partial u \partial r}} \; + \; {{n - 2}
\over r}
\; {{\partial \phi } \over {\partial u}} \; - \; {r^{ 2 - n}} \; {\partial \over {\partial
r}} \; \left ( {r^{ n - 2}} \, {\bar g} \; {{\partial \phi } \over {\partial
r}} \right ) = 0 \; \; \; .
\end{equation}

In principle, equations (\ref{gsol}-\ref{wave}) could be used for a
numerical treatment of the Einstein-scalar equations.  Given $\phi$ on an initial
light cone, equations (\ref{gsol}) and (\ref{gbarsol}) could be used to find the
metric on that light cone.  Equation (\ref{wave}) could then be used to evolve
$\phi$ to a nearby light cone.  However, in 4 spacetime dimensions,
Christodoulou\cite{christodoulou} finds that a change of variables gives rise to a 
nicer form of the wave equation.  
Define the operator
$ D $ by $ D \equiv {e^\nu} {n^a} {\nabla _a} $.
Then $ D $ is a derivative along ingoing light rays and we have
\begin{equation}
\label{Ddef}
D = {\partial \over {\partial u}} \; - \; {1 \over 2} \; {\bar g} \;
{\partial
\over {\partial r}} \; \; \; .
\end{equation}
Now in 4 spacetime dimensions define
$ h \equiv ({\partial / {\partial r}})  ( r \phi )$.
Then the wave equation (\ref{wave}) becomes
\begin{equation}
\label{wave4}
D \, h = {1 \over {2 r}} \; ( g \, -
\, {\bar g} ) \, ( h \, - \, {\bar h} ) 
\end{equation}
where ${\bar h} \equiv {r^{-1}} {\int _0 ^r} h \, dr$.
For the purposes of numerical simulations, the nice property of equation (\ref{wave4}) is 
that it involves a derivative only in the direction of 
ingoing  light rays. 
Another nice feature of equation (\ref{wave4}) is that in the 
absence of gravity 
($g={\bar g}=1$), the right hand side vanishes.  This is related 
to the fact that 
in general, the right hand side vanishes like $r^2$ as $r \to 0$. 
(The metic is smooth,
and therefore in a neighborhood of the origin it behaves like
a flat metric to some order in $r$).
This property is important for the following reason: spherically
symmetric metrics have coordinate singularities at the origin.  These coordinate
singularities are reflected in the appearance of inverse powers of $r$ in equation (\ref{wave}),
which can lead to  instabilities or inaccuracies in a numerical simulation.

These considerations suggest that instead of a numerical simulation of equation
(\ref{wave}) in $n$ dimensions, we should instead search for a new variable $h$
that satisfies $D h = 0$ in the absence of gravity.  Note that a quantity that 
satisfies $Dh = 0$ is constant along ingoing light rays and therefore satisfies
Huygens' principle.  However, it is well known that solutions 
of the wave equation
in flat spacetime satisfy Huygens' principle only in even 
spacetime dimensions.\cite{huygens} 
Therefore, we should only expect to find an appropriate new variable in even
spacetime dimensions.  Let $n$ be even and define $m \equiv (n-2)/2$.  For any
$\phi$ define 
\begin{equation}
\label{hdef}
h \equiv {{(m-1)!}\over {(2 m - 1)!}} \; {r^{1 - m}} \; {{\left ( {\partial \over
{\partial r}} \right ) }^m} \; \left ( {r^{2 m - 1}} \, \phi \right ) \; \; \; .
\end{equation}
(Here, the numerical factor is chosen for later convenience).  If $\phi$ is a
solution of the flat space wave equation (equation (\ref{wave}) with ${\bar g} = 1$),
then $h$ satisfies $D h = 0$.  That is, in the absence of gravity,
a solution of the wave equation $\phi$ gives rise to an $h$ that is constant along
ingoing light rays.  One can demonstrate this property of $h$ by using equation
(\ref{wave}) (with ${\bar g}=1$) and mathematical induction on $m$.

The advantage of using $h$ as the basic variable is that it 
tends to make computer
simulations more stable and accurate, especially near the 
origin.  The disadvantages
are (1) the method only works in even dimensions, (2) a 
seperate computer code must
be written for each value of $n$ and (3) as $n$ gets larger, 
the equations become
more complicated.  We expect that some numerical technique 
can be used to evolve
the Einstein-scalar equations in $n$ dimensions with a single 
code, with $n$ as a
free parameter, stably and with enough accuracy to treat critical 
gravitational 
collapse.  However, we have been unable to devise such 
a technique.  Therefore, 
in this work we use $h$ as our basic variable.  Due to the 
increase in complication
with increasing $n$, we treat only the case of $6$ spacetime dimensions.

We now specialize to the case of 6 dimensions.  The variable $h$ is given by
\begin{equation}
\label{h6}
h \equiv {1 \over {6 r}} \; {{\partial ^2} \over {\partial {r^2}}} \; \left
( {r^3} \, \phi \right ) \; \; \; .
\end{equation}
We must now use equation (\ref{gsol}) to express $g$ in terms of $h$ and integrals
involving $h$.  Define the quantities $s$ and $\bar s$ by
\begin{equation}
\label{sdef}
s \equiv {2 \over {r^2}} \; {\int _0 ^r} \; \hat r \; h(\hat r) \; d \hat r \; \; \; ,
\end{equation}
\begin{equation}
\label{sbardef}
{\bar s} \equiv {3 \over {r^3}} \; {\int _0 ^r} \; {\hat r^2} \; s(\hat r) \; d \hat r \; \; \; .
\end{equation}
Then, using equations (\ref{h6}-\ref{sbardef}) and (\ref{gsol}) we find
\begin{equation}
\label{g6}
g = \exp \left [ 18 \, \pi \; {\int _0 ^r} \; {{d \hat r} \over {\hat r}} \; {{\left (
s \, - \, {\bar s} \right ) }^2} \right ] \; \; \; .
\end{equation}
Equation (\ref{gbarsol}) specialized to 6 spacetime dimensions becomes
\begin{equation}
\label{gbar6}
{\bar g} = {3 \over {r^3}} \; {\int _0 ^r} \; {\hat r^2} \; g(\hat r) \; d \hat r \; \; \; .
\end{equation}
Using equations (\ref{h6}-\ref{gbar6}) in equation
(\ref{wave}) we find that the wave equation for $\phi $ in 6 spacetime dimensions becomes
\begin{equation}
\label{wave6}
D h = {3 \over r} \; ( g \, - \, {\bar g} ) \, ( h \, + \, {\bar s} \, - \, 2 s ) \; + \;
{{27 \pi} \over 2} \; {g \over r} \; {{(s \, - \, {\bar s} )}^3} \; \; \; .
\end{equation}
Equations (\ref{h6}-\ref{wave6}) are the full set of equations that are evolved in 6
spacetime dimensions.

\section{Numerical method}

The numerical method used is that of Garfinkle\cite{garfinkle}.  This method is 
based on numerical work of Goldwirth and Piran\cite{piran}, which is in turn
based on the analytical work of Christodoulou\cite{christodoulou}.  The spatial
grid is a set of points on an outgoing null cone, and each spatial grid point
is evolved along an ingoing light ray.  Given $h$ on a light cone, the code 
performs several integrals in turn: equations (\ref{sdef}-\ref{gbar6}).  
Near the origin, these integrals
are approximated by a Taylor series and evaluated using the 
slope of $h$ at $r=0$.  At all other points,
the integrals are evaluated using Simpson's rule.  

For each time step the quantity $h$ is evolved using equation (\ref{wave6}) 
and the quantity $r$ is evolved using
\begin{equation}
D r = - \; {{\bar g}\over 2} \; \; \; .
\end{equation}
These are essentially a set of uncoupled ODEs, one for each grid point.  
In the scheme of\cite{garfinkle} as the evolution procedes grid points
that pass through $r=0$ are lost.  
When half of the grid points are lost,
they are put back, interpolated between the remaining grid points.  
The critical solution is found by a search of a $1$-parameter space 
of evolved data to find the
boundary between those data that form black holes and those that do not.  
The outermost gridpoint is chosen to be the light ray that hits 
the singularity of
the critical solution.  This choice maintains resolution 
throughout the evolution.

\section{Results}

All runs were done with 300 spatial grid points.  The code was run in quadruple
precision on Dec alpha workstations and in double precision on a Cray YMP8. 
The initial data for the scalar field was chosen to be of the form
\begin{equation}
\label{initphi}
\phi (0,r) = p \; {r^2} \; \exp \left [ - \; {{(r - {r_0})}^2}/{\sigma ^2}\right ]
\; \; \; .
\end{equation}
Here, $p$ is our parameter, and $r_0$ and $\sigma$ are constants.
The critical solution was found to have discrete
self-similarity.  Let $u*$ be the value of $u$ at 
the origin at the singularity of the
critical solution.  Define the coordinates $T \equiv - \ln (u* - u)$ and 
$R \equiv r {e^T}$.  Then discrete self similarity means that $h$,
considered as 
a function of $T$ and $R$, is periodic in $T$.  
Figure 1 shows $h$ plotted
as a function of $T$ and $R$.  
Note that after some initial
transient behavior, $h$ becomes periodic.  To examine the 
periodicity more closely,
in Figure 2 we plot $h$ {\it vs} $R$ at two different times 
where the minimum 
of $h$ occurs at $r=0$.  Note that the curves agree, demonstrating 
periodicity
of $h$.  The period of $h$ is $\Delta = 3.03$.  Thus, in each period 
a typical length of
the system shrinks by a factor of ${e^\Delta} \approx 21 $.  
For comparison, recall that for $n=4$, $\Delta = 3.445$.

To compute the scaling exponent $\gamma$ we use the method of 
reference\cite{meandcomer}.
We evolve data for a range of parameters below the threshold of 
black hole formation.  For
each evolution we find the maximum of the absolute value of the 
scalar curvature on the
world line of the central observer
$R_{\rm max}$.  Plotting $\ln {R_{\rm max}}$ {\it vs} $\ln (p*-p)$ the result
is a straight line with a periodic wiggle, where the slope of the 
line is $- \, 2 \gamma$.  We use 50 values of $p$ equally spaced 
in $\ln (p*-p)$.  Figure 3 is a plot
of  $\ln {R_{\rm max}}$ {\it vs} $\ln (p*-p)$ along with the best
straight line fit to the points.  
The slope allows us to find $\gamma$.   
The result is $\gamma = 0.424$, so $M_{BH} \propto (p-p*)^{1.27}$.  
Recall that for $n=4$, $\gamma = 0.374$.  Note that $\gamma$ can also
be found by evolving data above the threshold of black hole formation
and plotting $\ln {M_{BH}}$ {\it vs} $\ln (p-p*)$.  We have done this
and the result is a straight line with a periodic wiggle where the slope
of the line is $\approx 1.27$.  However, our method gives a more
accurate treatment of subcritical collapse than of supercritical
collapse.  For that reason, we have used subcritical collapse to 
calculate $\gamma$. 

Comparison of our results for 6 dimensions with Choptuik's results for 4 
dimensions\cite{choptuik} seems to indicate that $\Delta$ is a decreasing
function of the dimensionality $n$ of spacetime.  The quantity $\gamma$ seems
to be an increasing function of $n$. It would be interesting to obtain more
information about these functions by studying critical collapse for other 
values of $n$.

\acknowledgments
 This work was partially supported by NSF grant PHY-9722039 to 
Oakland University.  We acknowledge the use of the Ohio Supercomputer
Center, where some of the computations were performed.

\begin{figure}[bth]
\begin{center}
\makebox[4in]{\psfig{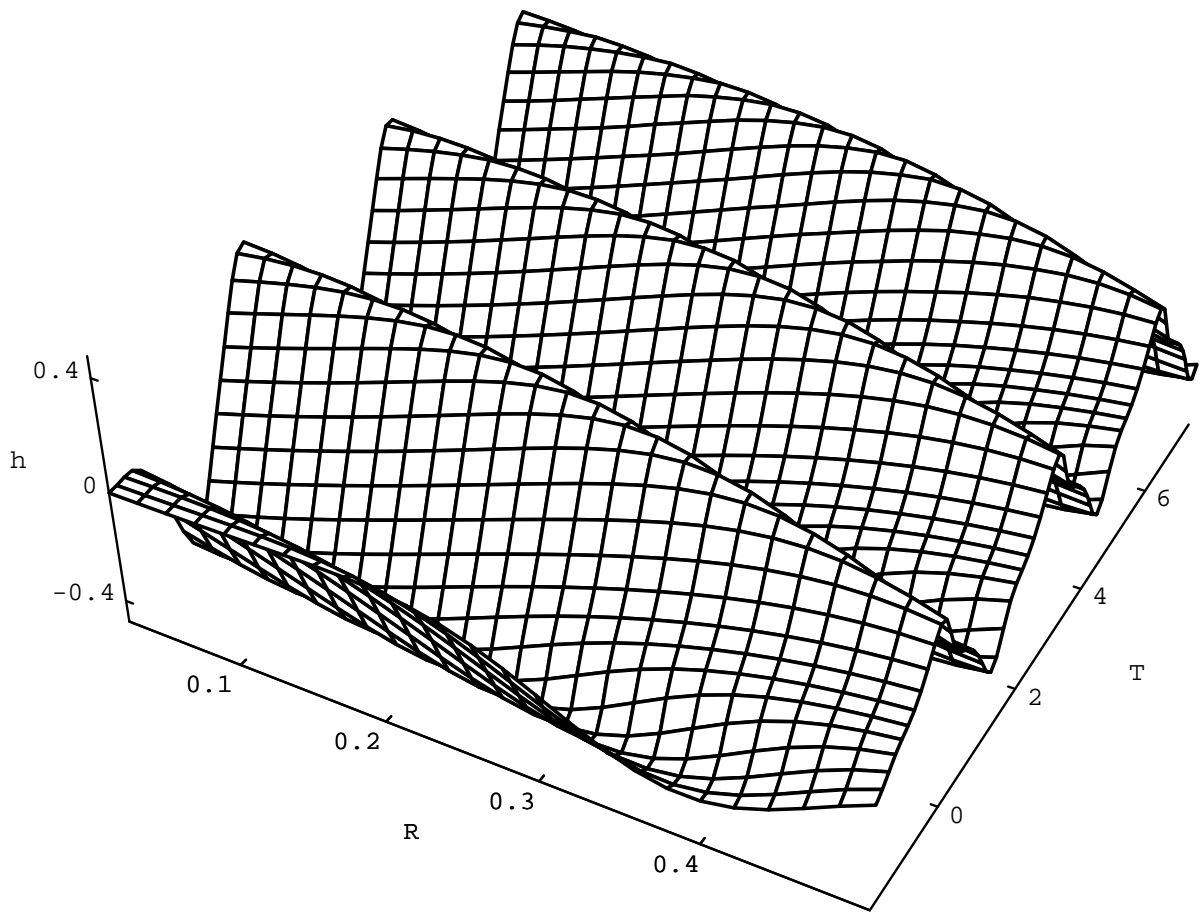}}
\caption{$h$ plotted vs. the rescaled coordinates $R$ and $T$.}
\label{fig1}
\end{center}
\end{figure}
\vfill\eject 

\begin{figure}[bth]
\begin{center}
\makebox[4in]{\psfig{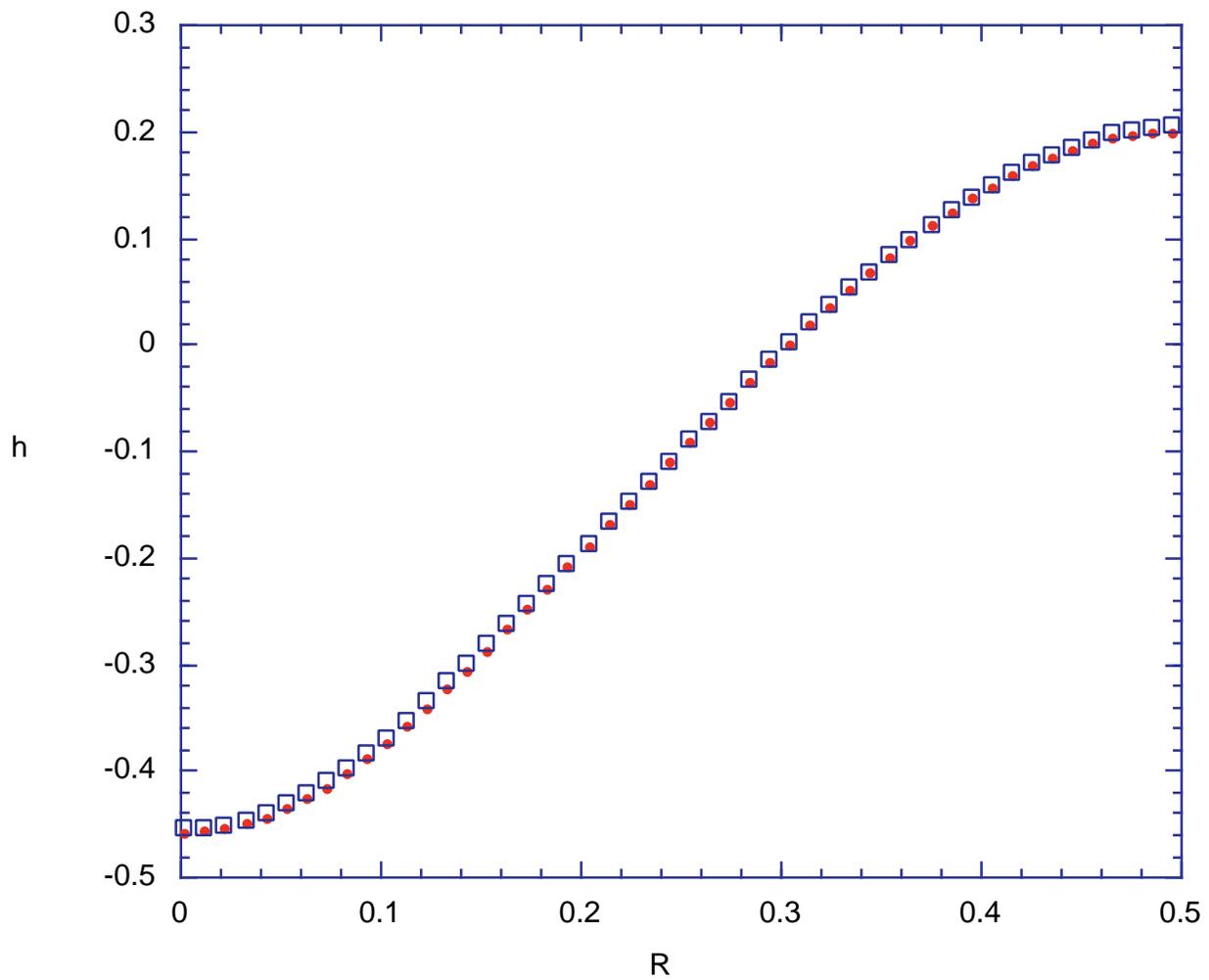}}
\caption{$h$ plotted vs. $R$ at two times when the minimum of $h$ is
at $r=0$.  Note that the two curves agree.}
\label{fig2}
\end{center}
\end{figure}
\vfill\eject

\begin{figure}[bth]
\begin{center}
\makebox[4in]{\psfig{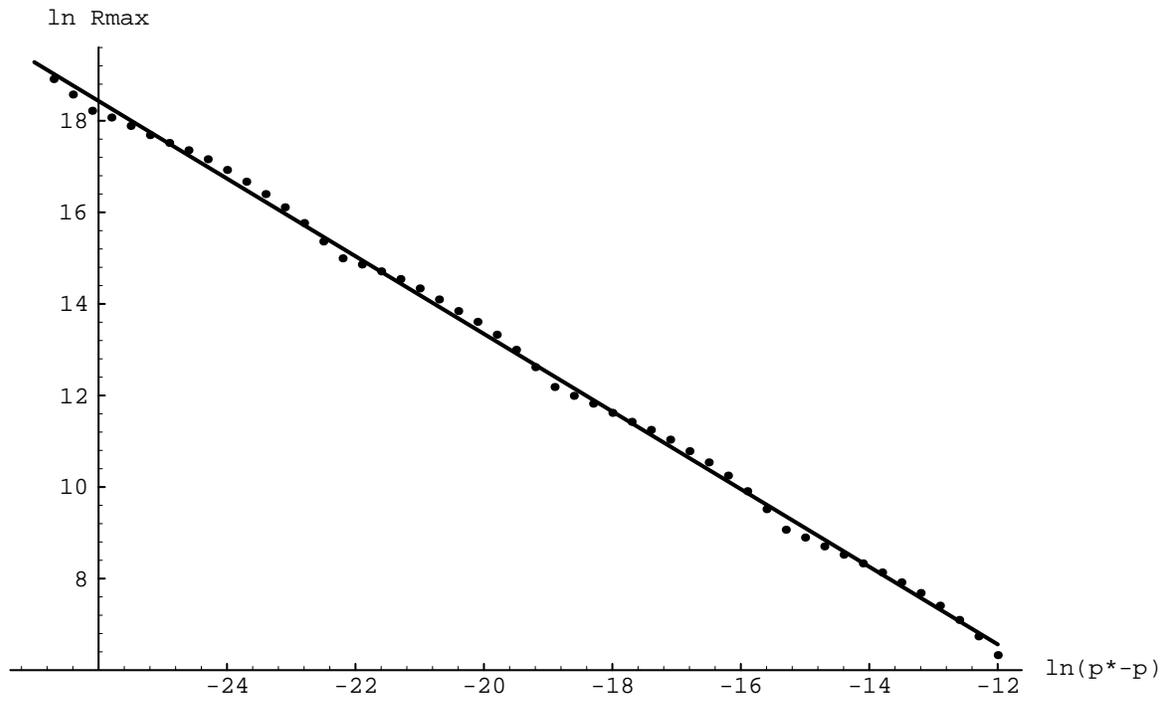}}
\caption{$\ln {R_{\rm max}}$ plotted vs. $\ln (p*-p)$ along with
the best straight line fit.  The curve is a straight line with a
periodic wiggle and the slope of the line is $- 2 \gamma $.}
\label{fig3}
\end{center}
\end{figure}

\end{document}